\documentclass[sigconf]{acmart}
\AtBeginDocument{%
  \providecommand\BibTeX{{%
    \normalfont B\kern-0.5em{\scshape i\kern-0.25em b}\kern-0.8em\TeX}}}

\copyrightyear{2024}
\acmYear{2024}
\setcopyright{acmlicensed}\acmConference[SIGCSE 2024]{Proceedings of the 55th ACM Technical Symposium on Computer Science Education V. 1}{March 20--23, 2024}{Portland, OR, USA}
\acmBooktitle{Proceedings of the 55th ACM Technical Symposium on Computer Science Education V. 1 (SIGCSE 2024), March 20--23, 2024, Portland, OR, USA}
\acmPrice{15.00}
\acmDOI{10.1145/3626252.3630899}
\acmISBN{979-8-4007-0423-9/24/03}

\begin{document}

\title[Not Just Training, Also Testing]{
Not Just Training, Also Testing: High School Youths’ Perspective-Taking through Peer Testing Machine Learning-Powered Applications 
}

\author{Luis Morales-Navarro}
\email{luismn@upenn.edu}
\orcid{0000-0002-8777-2374}
\affiliation{%
  \institution{University of Pennsylvania}
  \city{Philadelphia}
  \state{PA}
  \country{United States}
}

\author{Meghan Shah}
\email{megshah@upenn.edu}
\orcid{0009-0006-1508-3031}
\affiliation{%
  \institution{University of Pennsylvania}
  \city{Philadelphia}
  \state{PA}
  \country{United States}
}

\author{Yasmin B. Kafai}
\email{kafai@upenn.edu}
\orcid{0000-0003-4018-0491}
\affiliation{%
  \institution{University of Pennsylvania}
  \city{Philadelphia}
  \state{PA}
  \country{United States}
}


\renewcommand{\shortauthors}{Luis Morales-Navarro, Meghan Shah, \& Yasmin B. Kafai}
%% use of ampersand (\&) versus ''and'' is to saves space.

\begin{abstract}
  Most attention in K-12 artificial intelligence and machine learning (AI/ML) education has been given to having youths train models, with much  less attention to the equally important testing of models when creating machine learning applications. Testing ML applications allows for the evaluation of models against predictions and can help creators of applications identify and address failure and edge cases that could negatively impact user experiences. We investigate how testing each other's projects supported youths to take perspective about functionality, performance, and potential issues in their own projects. We analyzed testing worksheets, audio and video recordings collected during a two week workshop in which 11 high school youths created physical computing projects that included (audio, pose, and image) ML classifiers. We found that through peer-testing youths reflected on the size of their training datasets, the diversity of their training data, the design of their classes and the contexts in which they produced training data. We discuss future directions for research on peer-testing in AI/ML education and current limitations for these kinds of activities. 
\end{abstract}





\maketitle

\section{Introduction}
Recent calls to increase artificial intelligence (AI) literacy in K–12 education place an emphasis on the importance of engaging young people with big ideas, preparing  them to be critical technology users and designers, and highlighting the role of AI/ML in different career paths \cite{diPaola2022artificial, touretzky2022artificial}. Fostering an understanding of machine learning (ML), which entails using data instead of code to influence the behavior of algorithmic systems, is essential for AI literacy \cite{long2020ai, zimmermann2019youth, tedre2021ct}. ML provides rich opportunities for K-12 youths to experiment with training data and learning algorithms. Yet, while there are numerous studies that have investigated youths’ understanding of machine learning, most have done so from the perspective of having them train models \cite{sanusi2023systematic}, only few studies (e.g., \cite{druga2021children, vartiainen2021machine, tseng2023collaborative}) have examined what youths can learn by also testing the models they trained.  Furthermore, these studies highlight the need for providing scaffolds to support youths in testing their ML models.

In this paper we review the role of testing in learning to build ML applications and share findings from a two-week long summer workshop with high school youths (ages 13-15) in which they built ML-powered physical computing projects. In our analysis we addressed the following research question:  How did testing each other's projects support youths to take perspective about functionality, performance, and potential issues in their own projects? Findings show how testing may support youths in reflecting and making sense of how data diversity, class design, the size of their data sets, and the context of training data may affect model performance. Finally, we discuss challenges and opportunities to further incorporate testing in ML education.

\section{Background}

Tedre and colleagues \cite{tedre2021ct} contend that incorporating ML into computing education challenges some of the computational thinking practices that have been adopted over the past decade. The case of testing and debugging is a particularly relevant because to “locate a structural defect in a program does not apply when the internal weight matrix of a neural network misclassifies [an input]” \cite[p. 2]{tedre2021ct}. Instead, they propose that testing involves evaluating the model against predictions \cite{tedre2021ct}. Indeed, as Shapiro and colleagues \cite{shapiro2018machine} argue ML models are opaque and not human-readable algorithms where there are proofs of correctness. Contrary to the traditional computing education paradigms, ML is an empirical science \cite{langley1988machine} that requires experimentation, formulating hypotheses and evaluating models against predictions \cite{shapiro2019new}. In professional practice, testing has become increasingly important to ensure that ML models work as expected, detect edge and failure cases and identify potential harmful biases \cite{zhang2020machine}. Yet, current ML education efforts center on having youths train models with little attention given to how they test models and how testing can support students in debugging, identifying issues and making sense of their models' functionality. In this section, we discuss the role of testing in ML, current efforts in ML education, and implications for future research. 

Over the past 15 years testing has received increasing attention in ML research and practice as perspectives have shifted from approaching models as “untestable” to developing heuristics for ML testing, that is  “detecting differences between existing and required behaviors of machine learning systems” \cite[p. 2]{zhang2020machine}. These efforts have been driven from desires to make transparent the functionality and performance of such systems as well as to mitigate harmful biases that may disproportionately affect historically marginalized groups. While testing in computing has traditionally centered on detecting errors in code; in ML, testing may help detect issues in the training data, the learning algorithm or the framework being used which all are key components of the process of building models \cite{zhang2020machine}. Furthermore, with concerns about algorithmic injustice in AI/ML systems, evaluating models has become more imperative through both testing and auditing (which involves systematically evaluating models from the outside in) \cite{metaxa2021auditing}.	
 
However, in K-12 computing education most efforts have centered on having youths train models with less attention given to testing. For instance, recent systematic reviews rarely mention studies that discuss testing in teaching, learning, and assessing ML in K-12 contexts \cite{sanusi2023systematic, rauber2022assessing, druga2022landscape, marques2020teaching}. This is not only an issue in K-12 education but also in undergraduate ML education where testing is not always included as a necessary step for learners to consider in the ML pipeline \cite{fiebrink2019machine, schapire_2014}.
	
 Some research studies have brought up testing as part of learning activities but provide little details on how youths test models and what they can learn from testing \cite{williams2021teacher,voulgari2021learn,lee2022preparing,kaspersen2022high,hitron2019can,burhans2017arty,akram2022towards}. Some of these studies claim that testing is crucial for youths to understand model behavior, build hypotheses and conduct experiments to make observations about how the systems work \cite{ng2022using}. 

The first studies that engage K-12 youths in testing ML applications showcase some interesting trends. For instance, Vartiainen \cite{vartiainen2020learning, vartiainen2021machine} and colleagues argue that when building models for personally relevant projects, iteratively testing can support children (ages 3-9 and 12-13) to reason about how and when models break and to build explanations and hypotheses about how the models could be improved. Similarly, in a study conducted with primary and middle school students children (ages 7-13), Dwivedi and colleagues \cite{dwivedi2021exploring} found evidence that testing supported youths in coming up with hypotheses and explanations for model behavior, particularly making connections between the qualities of the training data (data diversity) and model behavior when exposed to new data. In this study participants were grouped in pairs, they trained models individually and swapped models with their partners to test them.

Furthermore, Druga and colleagues \cite{druga2021children} argue that when training and testing models children  (ages 7-12) “engage in the scientific method by formulating hypotheses about machine intelligence, then generate scenarios for testing, and finally refining their understanding either by affirming their initial hypotheses or formulating new ones.” In this study participants interacted with a platform that integrated model training, block programming, and live-testing. Their analysis suggested that children come up with edge cases and common cases to test the “abilities” of the systems they interact with, making inferences about why the models may fail and coming up with similar examples to furhter test their inferences. Additionally, a more recent study by Tseng and colleagues \cite{tseng2023collaborative} introduces collaborative model building with a tool that enables multiple learners to create training and testing data sets, train and test models. In this study children (ages 11-14) and their parents were guided through the process of building models. Here iteratively and collaboratively testing models supported participants in identifying issues of data diversity, class imbalance, and data quality. It is worth noting that in this study participants did not only engage with a live testing interface but could also create testing datasets over time to see how model performance changed. 

However, Zimmermann-Niefield and colleagues \cite{zimmermann2019youth} show that the availability of live model testing tools is not enough to support youths to understand the iterative nature of building models. In a case study, they observed how a student, after finding that a model that classified movements did not perform as expected, decided to change the testing movements instead of trying to fix the model. This suggests that providing youths with tools for testing is not enough and scaffolding may be necessary for learners to fully benefit from iteratively testing and training their models. In a separate study, Zimmermann-Niefield et al. \cite{zimmermann2020youth} also identified that participants rarely tested their models on their own, only iterating on testing and training when prompted by researchers. They found that often youths re-tested their models repeatedly without making any changes to them, voicing the idea that solely testing (without making changes to the training data and retraining) could improve model performance. These two studies illustrate the importance of scaffolding testing approaches to support learners.

In this study we investigate how “peer testing”  or collaboratively testing models with peers may support learners to reflect through perspective-taking, identify failures in their own models and come up with ideas to improve model performance. Here, we build on Ackermann’s \cite{ackermann2012perspective} conceptualization of perspective-taking as “people’s ability to experience and describe the presentation of an object or display from different vantage points,” [p. 28] or the process in which learners move away from their own standpoint, “to take on another person's view” [p. 29]. She argues that this process is necessary for learners to construct deeper understandings of how the applications they create work. Ackermann, like Kegan \cite{kegan1982evolving}, claims that deeper understanding is achieved by both “diving in,” actively engaging in addressing a problem, and “stepping out” to reflect and take perspective on the task at hand. This framework of perspective-taking has been widely used in computing education and learning sciences research to investigate various topics including program comprehension \cite{strawhacker2015want}, understanding in building simulations of complex systems \cite{wilensky2006thinking}, and representations in data-science \cite{kahn2022learning,roberts2022examining}. 

Through the lens of perspective taking, we engaged high school youths in peer testing during a two-week workshop in which they created ML classifiers for ML-powered electronic textile (e-textile) physical computing projects. Electronic textiles involve using microcontrollers, sensors, actuators, and sewn circuits to create physical computing projects \cite{buechley2006construction}. During the workshop youths created personally relevant projects, iteratively training and testing ML classifiers, designing e-textiles, sending the outputs of the classifiers (via serial communication) to the microcontrollers connected to the e-textile projects. We focus in our analysis on model testing sessions to examine how testing each other's projects supported youths to take perspective about functionality, performance, and potential issues in their own projects.

\begin{figure*}[h]
  \centering
  \includegraphics[width=\linewidth]{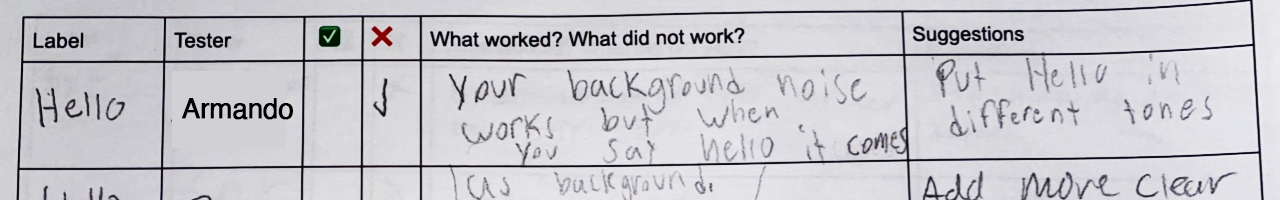}
  \caption{Worksheet in which youths documented their testing and provided suggestions.}
  \Description{}
  \label{fig:sheet}
\end{figure*}

\section{Methods}

\subsection{Context}
We conducted a two-week-long in-person workshop at a science center located in the Northeastern United States with 19 youths (ages 13-15 years) (11 with guardian consent and their own assent to take part in research) that had demonstrated an interest in STEM by participating in an out of school program designed to deepen participation for historically excluded communities. The 11 consenting youths that had taken part in the science center program for at least one year and thus already knew each other. Four participants self-identified as female, two as non-binary, and five as male. Seven participants identified as Black, two identified as Asian, one as Latinx and one as North African. Nine youths had taken computer science courses at school or workshops outside of school, yet none had taken workshops or courses on AI/ML. The organizers of the program invited youths to participate in the study via email and through paper handouts. Parents received consent forms prior to the study (which included a brief explanation of the research) and youths assented to their participation. The study protocol was approved by the University of Pennsylvania’s institutional review board.

During the workshop youths learned about e-textiles and ML classifiers. Each workshop session lasted 3.5 hours including a daily 30 minute community-building activity and a 15 minute snack break. During the first week of the workshop youths engaged in structured activities to learn about e-textiles, machine learning classifiers, and creating e-textiles projects that integrate machine learning classifiers. The hands-on project-based e-textile activities supported youths to learn about programming the micro:bit microcontroller, sensors, actuators, circuits and sewing with conductive thread. These were based on the Exploring Computer Science electronic textiles curriculum \cite{kafai2019stitching}. Following hands-on activities engaged youths in learning about AI/ML, different types of ML (supervised, unsupervised, generative), the ML pipeline \cite{fiebrink2019machine}, and training and testing image, audio, and pose classifiers created using first Google’s Teachable Machine \cite{carney2020teachable} and later \href{https://ml5js.org/}{ml5.js} (a JavaScript high level interface to TensorFlow designed for beginners). Following, they used serial communication to send the outputs of their classifiers to the micro:bit. During week two participants spent four days building personally relevant projects that incorporated both electronic textiles and ML.
 
Testing sessions using live classification played an important role during both weeks. In the first week, participants tested each other’s models on two occasions. During the following week, two days were structured so that youths could work on their projects in the first part of the day and spend the second part of the day testing. Testing was driven by the creators, who decided what to test, invited their peers to test the models and documented their testing in a worksheet (see Figure \ref{fig:sheet}). We intentionally scheduled these testing sessions to make sure that all participants tested their models and to build an environment where testing was recognized as an important aspect of model building.

\subsection{Data Collection \& Analysis}

We video and audio recorded participants while testing each others’ projects. We used an automated transcription tool to process audio recordings and reviewed the transcripts to ensure their accuracy. We also collected participants’ artifacts and testing sheets. Our analysis is grounded in traditions of qualitative and learning sciences research in computing education \cite{tenenberg2019qualitative,margulieux2019learning}. We conducted two rounds of analysis to identify common themes across the data \cite{braun2012thematic}. In a first round of analysis, two researchers inductively coded the transcripts in order to create an initial coding scheme. We clustered the themes into a codebook with categories for instances in which youths talked about data diversity, class design, context of data collection, and dataset size in relation to testing. The codebook was then applied to all transcripts and test sheets in a second round of analysis by the same two researchers. The researchers engaged in dialogue with the data while they coded jointly, seeking consensus and iteratively addressing disagreements with a third researcher familiar with the data. Because this is an exploratory study, with a small number of participants, we prioritized establishing unanimous agreement on all coding (by coding collectively and resolving our divergent opinions through extensive discussion) over reliability (when coders apply the same scheme independently on the same data) \cite{mcdonald2019reliability}. All names used in this paper are pseudonyms.

\section{Findings}

From the recordings and testing sheets, in which youths documented what they tested, we identified 50 instances in which youths found failure cases in their peers' projects, that is cases in which testers expected a different outcome than what was provided by the ML classifiers.  Below we describe in more detail how youths, through perspective taking, were able to identify failure cases in their peers' projects and how they reflected about these with regards to their own projects.

\subsection{Taking perspective on data diversity}

\begin{figure}[h]
  \centering
  \includegraphics[width=\linewidth]{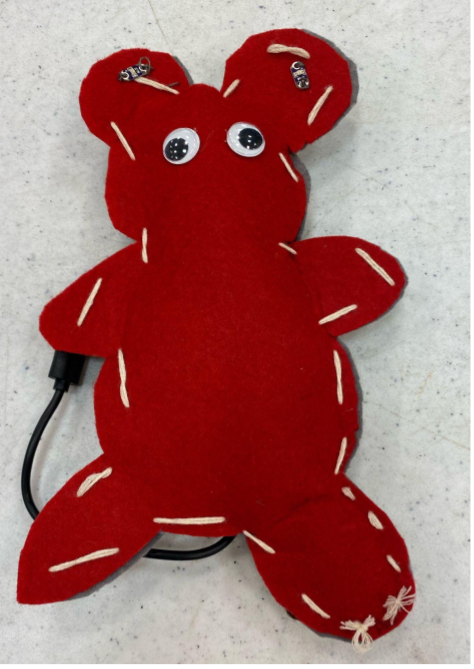}
  \caption{A plushy for children that identified key words and played melodies (for example if it identified the word “twinkle” it would play Twinkle, Twinkle, Little Star).}
  \Description{A plushy toy made of felt}
  \label{fig:plush}
\end{figure}

When testing each other’s projects youths identified several failure cases which they ascribed to issues of data diversity. Of notice,  26\% of the instances in which youths identified failure cases were related to issues of data diversity. For example, Armando and Jamal had their peers test an audio classifier they created for their final project, a plushy for children that identified key words and played melodies (e.g., if it identified the word “twinkle” it would play \textit{Twinkle, Twinkle, Little Star}) (see Figure \ref{fig:plush}). Aashvi, after testing the project, found the word “row” was misclassified most of the times, she described this behavior saying “It seems confused on what it hears” recommending that his peer “[added] more data with different voices.” Three different testers came to the same findings and recommendations for Armando and Jamal. In Dwayne’s project (also a sound classifier) testers also recommended adding more diverse samples with Rafik suggesting using different accents and different ways of saying words.

\begin{figure}[h]
  \centering
  \includegraphics[width=\linewidth]{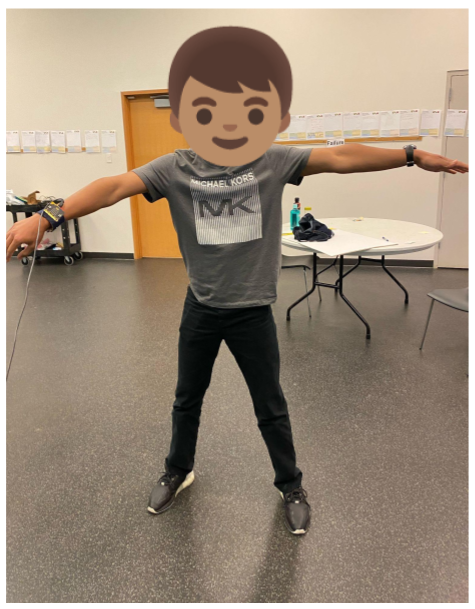}
  \caption{youths testing a game that prompted users with a dance move on a LED screen on a bracelet, if users made such a move they received a point and were prompted with a new pose. Otherwise, for every wrong move users lost a point.}
  \Description{A teenager makes a tree pose while having a bracelet with a microcontroller.}
  \label{fig:test1}
\end{figure}

After identifying failure cases in their peer’s projects and making suggestions on how these could be improved, youths reflected on their own projects and what they learned from their peers' testing. Keesha, who was working on a game that recognized different dance moves (see Figure \ref{fig:test1}) realized that when creating pose classifiers “It’s better to have different types of people height wise, look wise, and even you know baggy clothes” in the training dataset to address some of the concerns of her peers. Similarly  after seeing their peers testing his project, Tavon, who built a pose classifier of different jutsu (moves or skills) used in Naruto (a popular anime series), reflected: “it's better to use different types of body types because of the way that the program recognizes bodies.” Aashvi added that not only including different bodies in the training data was important but also considering how the data was gathered and in particular the angle of the camera: “You never know what the camera angle is going to be exactly so... using different camera angles may help it much later on when doing a prediction” she said. These examples illustrate how perspective taking supported youth, first by distancing themselves from their own projects to identify issues of data diversity in their peers’ classifiers and later by reflecting on their own projects based on the feedback provided by their peers. 

\subsection{Taking perspective on the number of samples used to train the models}

When testing each other’s projects youths identified several failure cases which they ascribed to issues related to the amount of data used to train the models. Issues related to sample sizes were present in 26\% of the instances in which youths identified failure cases. Jamal for example, when testing Naeem’s audio classifier, realized that it had only 10 samples for each class and hypothesized that this was the reason why the model “works sometimes but other times doesn't,” he recommended increasing the size of the dataset explaining “you need more data so it could tell which is which.” Youths did not just argue for having larger training datasets but considered possible differences in the amount of data used for each class and how this may create biases. Whereas they did not mention class balance issues explicitly they discussed these in their own words. When testing Armando and Jamal’s project, Rafik noticed that the audio model would classify random background noise as key words (the project had the classes “row”, “twinkle,” “ball,” and “background”). He ascribed this issue to the fact that there were only 4 samples in the background noise class and more than 20 samples for each of the other classes.

After testing youths reflected on their own projects. Armando noted “a lot of suggestions we got were we need more examples [...] we need to increase the number of background samples.”  This shows how testing provided an opportunity for Armando to distance himself from his project, listen to his peers’ feedback and come up with concrete next steps. Reflecting on how his peers tested his audio classifier, Dwayne said “we've gotten a lot of good feedback on the background class” which included samples taken in different environments and had about three times as much data as any other class. “But our “hello” class, it’s small,” he continued. This shows that peer testing supported Dwayne to recognize that differences in class sizes in the training data may affect model performance.

\subsection{Taking perspective on the context of data production}

Testing each others’ projects enabled youths to think about the context in which the data was produced. Of the instances analyzed 22\% involved youths describing issues related to how the context of the data could affect the performance of their models. They talked about how background noise could affect audio classifiers, how the background of pictures and other elements such as hands could affect image classifiers, and how the camera angle used while creating training data could affect pose classifiers. For instance, when testing Jamal and Armando’s project, Naeem noticed that the model only worked when the room was quiet which led him to recommend adding data samples from other louder contexts. Trisha observed that in an image classifier designed by Dwayne wearing a face mask could confuse the model because in some of the training data for one of the classes Dwayne was using a face mask.

After the peer-testing activity youths reflected on how the context in which data was produced affected the performance of their own models. Armando reflected on his sound classifier: “the room, like this one right here, is like a little bit too loud for testing the project because we did like our sampling in the gardens.” Alicia, who had been working on an image classifier of recyclable materials (bottles, soda cans, paper) also noted “I learned that the lighting and the backgrounds of images affects the outcome.” Testing supported youths to reflect and consider possible spurious relationships in which the contextual aspects of data samples may affect the performance of their peers and their own models. 

\subsection{Taking perspective on class design}
Peer testing supported some youths to rethink the design of the classes in their classifiers. When testing each other's projects, youths ascribed some (6\%) of the issues they encountered to problems with the design of classes used in the classifiers. Rafik identified that the classes in Dwayne’s audio classifier were “too broad” (e.g., a class labeled “hello” included data samples of people saying “hi”, “hello,” and “greetings”) realizing that only “hi” was being correctly classified. He recommended that his peers instead created more classes that were more specific (i.e., “one for each word”). 

Testing provided an opportunity for learners to see how their peers decided on classes or categories for their classifiers and to reflect on their own classifiers. After participants tested Tavon’s justu project, Tavon realized that “poses for class one and class two, they were too similar” and they had to come up with a new pose that was different enough for the classifier to perform better. Armando also reflected on a similar issue explaining that the classifier “is confused if [two classes] look too similar.” Testing helped youths see how other people built classes, and how decisions around class design may impact model performance, which supported reflection on their own models.
 
\subsection{Taking perspective on testing}

After completing the testing activity, youths also reflected on the testing process and its importance in creating ML models. Naeem told his partner “testing often is a good thing, and more testing early.” He added, “[because] stuff that is important can't just be deployed without proper testing.” Jamar, in a conversation with Trisha, expressed a similar idea saying that “to make better models we have to revise it and test it often.” Trisha nodded, saying “also work together with other people to give different test examples;” recognizing the role peers can play in testing models.  Rafik added “test as often as possible because the more moving parts you have it's gonna be harder to find that part that has a bug.” These quotes illustrate how participants reflected on the testing activities and their value in creating ML models.

\section{Discussion}
In this study, we documented how testing sessions in which learners tested their peers' ML models may support perspective-taking on how data diversity, the amount of data, the context, and to a lesser extent the design of classes, in which data is produced may influence model performance. Similarly to previous studies \cite{dwivedi2021exploring, tseng2023collaborative}, we found evidence that testing supported learners to come up with hypotheses of how their models function and make connections between the design of training datasets and model performance. There are several aspects of our peer testing approach that deserve further discussion: (a) the scheduling of testing sessions, (b) the provision of testing prompts, and (c) the role of peers.

For one, building on previous studies that highlighted the importance of scaffolding testing \cite{zimmermann2019youth, zimmermann2020youth} we decided to conduct testing sessions. Our approach in which all students tested each other’s models at the same time scaffolded testing and made testing an integral part of the process of building models. Having a dedicated time in the classroom for iterative testing may be particularly helpful to support students’ perspective-taking. Second, we also provided prompts in the form of test worksheets that they used to keep track of their observations and findings. Finally, we engaged youths in discussion with their peers after testing to facilitate perspective taking. 
 
We found that testing sessions supported youths to experience and describe failure cases both as testers and creators of ML classifiers affording them “different vantage points” \cite{ackermann2012perspective} to make sense of ML models. Across all these different emerging themes we observed how youths engaged in what Ackermann \cite{ackermann2012perspective} calls “stepping out,” moving away from their own stand-points to identify moments in which their peers’ models failed and provide feedback. Later, when “diving in” into their own models they were able to reflect on what their peers found when testing, and come up with plausible next steps that may improve model performance. 
Our analysis also revealed that youths had different successes in recognizing some functionality and performance issues in their peers’ designed  machine learning applications. Issues of data diversity, the amount of data, and the context were more frequently recognized than the design of classes. This finding, of course, could also be due to the ML applications their peers had designed. But it also points towards potential challenges, that some data issues may be harder to understand and detect for youths than others, which requires further investigation of scaffolds that might support youths in this process. 

We recognize that the testing activities in our study were pretty low tech and based on peers’ experiences of live-testing models without involving any measures of overall accuracy and/or precision or the creation of large testing data sets. While more data driven testing activities may be beneficial, our study highlights how the social aspects of peer testing and taking “different vantage points” as testers and creators is valuable for students. Data-driven testing activities should also aim to create environments where students interact with their peers and take different roles. Here the work of Tseng and colleagues \cite{tseng2023collaborative, tseng2023co} creating collaborative testing datasets is particularly promising. 

There are several directions for future research. Studies could investigate different approaches that students may have to testing models, the differences between testing their own models and testing peer models, or even using testing activities as assessment scenarios. Furthermore, there are opportunities to connect testing activities to critical conversations about AI/ML systems and their potential harmful biases. Here, methods of algorithm auditing \cite{metaxa2021auditing} could be adopted for learners to evaluate models and everyday AI/ML applications from the outside in. It is worth noting that while previous studies and our study centered on detecting issues in training data, testing in ML also involves identifying issues in the learning algorithm or the framework being used which all are key components of the process of building models \cite{zhang2020machine}. Future work should also investigate how to support K-12 youths in testing ML systems more comprehensively.

\begin{acks}
With regards to Mia Shaw for support in data collection.
\end{acks}

\bibliographystyle{ACM-Reference-Format}
\bibliography{sample-base}

\end{document}